# Agent Based Distributed Control of Islanded Microgrid – Real-Time Cyber-Physical Implementation

Tung Lam NGUYEN[1,2], Quoc-Tuan TRAN[2], Raphael CAIRE[1], Catalin GAVRILUTA[1], Van Hoa NGUYEN[1]
[1]Grenoble Electrical Engineering Laboratory (G2ELAB), Grenoble, France
[2]CEA-INES, Le Bourget-du-lac, FRANCE
tung-lam.nguyen @g2elab.grenoble-inp.fr, quoctuan.tran@cea.fr, raphael.caire@g2elab.grenoble-inp.fr,
catalin.gavriluta@g2elab.grenoble-inp.fr, van-hoa.nguyen@grenoble-inp.fr

*Abstract*—In the hierarchical control of an islanded microgrid, secondary control could be centralized or distributed. The former control strategy has several disadvantages, such as single point of failure at the level of the central controller as well as high investment of communication infrastructure. In this paper a three-layer architecture of distributed control is given, including the device layer, the control layer as well as the agent layer. The agent layer is a multi-agent system in which each agent is in charge of a distributed generation unit. Due to communication network constraints, agents are connected only to nearby neighbors. However, by using consensus algorithms the agents can discover the required global information and compute new references for the control layer.
The proposed control system is tested on a microgrid scenario which includes paralleled inverter sources. For this, the system is implemented on a real-time cyber-physical test platform that combines real-time simulation models running in OPAL-RT with a network of ARM-based computers, representing the agents.

## I. INTRODUCTION

Microgrids (MGs) are considered to be one of the major changes required in order to reduce the carbon footprint and to offer autonomous energy provision as well as resilience in disaster relief. In general, a MG consists of a cluster of distributed generators (DGs), loads, energy storage systems and other equipment, which can operate in islanded mode or grid-connected, and can seamlessly transfer between these two modes [1].

A hierarchical structure comprised of primary, secondary and tertiary control is typically used to control MG. The primary control, typically droop-based, is designed to stabilize frequency and voltage by using only local measurements. It is necessary to have a fast response time in this control level in order to keep frequency and voltage near the nominal values. The secondary control, implemented in either centralized or distributed fashion, is responsible for the restoration of the frequency and voltage by compensating the deviations caused by the primary control. At the top level, tertiary control manages the power flow to the main grid and optimizes certain economic or operational aspects.

The comparison between centralized and distributed control has been properly discussed in [1]–[4]. Figure 1 illustrates the main differences between the two control approaches. The distinct feature of the distributed approach is that the information involved in the control algorithm is not global, but adjacent for any given unit. Also, the length of the communication links is often shorter, which offers better and more reliable latency. Moreover, the risk of overall system failure can be reduced, because the system does not depend on a sole central controller.

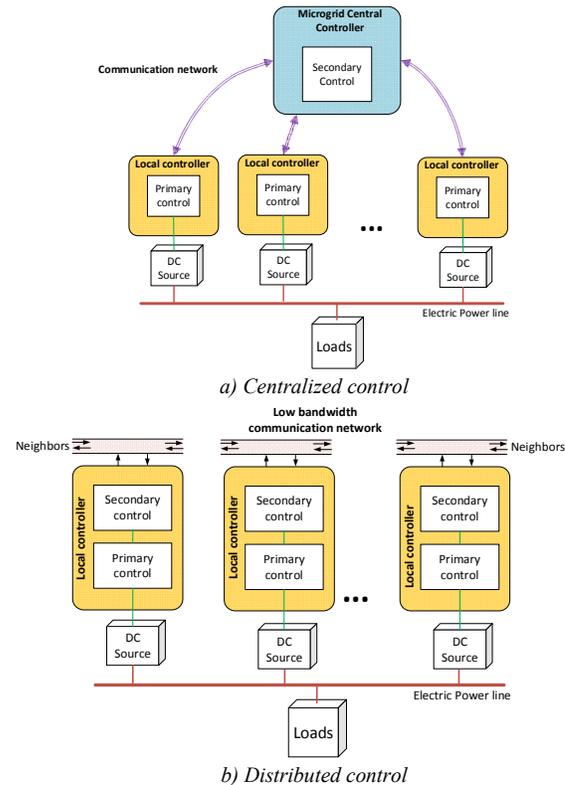

Figure 1. Control strategies in microgrid

In islanded mode, MG operation is more sensitive to frequency disturbance compared to the grid-connected case due to the lack of inertia in the system, the intermittency of renewable generators and varying demand of loads. DG controllers need to be properly coordinated to satisfy the

This work has been elaborated within the Erigrid project, supported by the H2020 Programme under Grant Agreement No. 654113, see https://erigrid.eu/ website for more information and also by Vietnamese government.

demand requirement and maintain a stable frequency. In this paper, a distributed control structure is proposed for implementing the frequency control of an islanded MG. This is achieved by sharing the active power demand among multiple inverter-based DGs. In order to provide a distributed control, the multi-agent system approach is proposed.

The given control architecture includes three layers: the device layer, the control layer and the agent layer. An experimental setup consisting of a real-time simulation model running on OPAL-RT and a real TCP/IP communication network are established in order to validate the operation of the proposed control strategy. The device layer and control layer are considered part of the physical process and are running on the real-time simulator, meanwhile the agent layer is created by a set of ARM-processors connected between them via TCP/IP.

The multi-agent system (MAS) is an innovative technology that has been recently used in a wide range of applications in power systems [5]. The agent based distributed control is also presented in [6], [7]. However, in these works, the inter-agent transmission latency which plays an important role in distributed control is neglected or simulated as a deterministic time. In this paper, a communication network with real and variable latencies is considered in the process of validating the proposed distributed control method. Each agent in our system is connected only to nearby neighbors due to communication constraints. Thus the consensus algorithm is implemented in order to get the global information, specifically the value of frequency deviations

The rest of the paper is organized as follows. Section II introduces the proposed layer structure of control in MG. The laboratory setup is presented in Section III. In section IV, experimental results are shown to validate the proposed method. Section VI concludes the paper and outlines possible future directions.

## II. LAYER STRUCTURE OF CONTROL IN MICROGRID

In this paper, a structure using a multi-agent system that leverages the consensus algorithm is proposed in the context of microgrid control. The distinct feature of this structure is that it takes into account the communication network, which is vital in the modern grid. The topology of the DG controllers is divided into three layers as depicted generally in Figure 2. The function of each layer will be described specifically in the case of distributed control strategy of islanded microgrids. Nevertheless, this structure is flexible, making it entirely possible to extend it to many other cases.

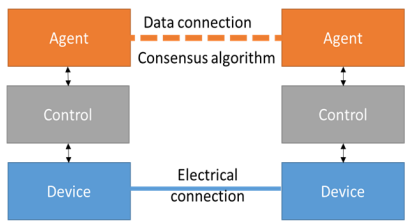

Figure 2. Layer control structure

A detailed representation of a DG controller in the proposed structure is illustrated in Figure 3.

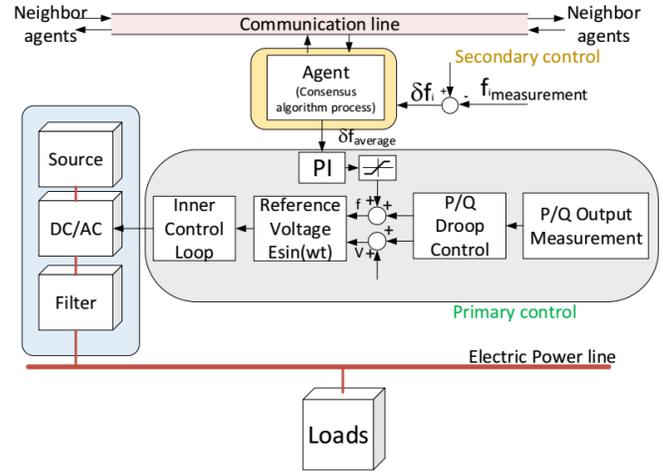

Figure 3. DG controller diagram

### A. Device layer:

This layer contains the physical components. Measurement devices send instantaneous signals to upper layers. The output voltage and current is sent to the Control layer for calculating active and reactive power as well as feedback values in the inner control loops. The frequency deviation measured at the DG's output is transferred to the Agent layer, which in turn forwards it to its neighbors. In return, power inverters will receive pulse signals from the controller that will adjust the output voltage so that it follows the calculated reference value.

When the MG operates in island mode, DG units are responsible for maintaining the voltage and the frequency values within operational limits. The intermittency of generation units such as PV or small wind turbines are often controlled following a maximum power point tracking algorithm (MPPT). Thus battery energy storage systems (BESSs) are used as backup power supply, as they can deal with supply-demand imbalance problems caused by the variation of renewable DGs and load demands. For the convenience of the investigation in this paper, BESS is simplified as the ideal DC supply. The MG with multiple inverter based sources operating in parallel will be investigated in this work. Each DG is a voltage controlled source controlled by a grid-forming inverter [8].

### B. Control layer:

This layer responds to changes in the system operation, and provides corresponding signals to components in the Device layer. In this paper, the Control layer is designed to resist to the instability of frequency and voltage amplitude in microgrids. When a change in the MG takes place due to the variation of DGs or loads, the primary control of the DG will react instantaneously and will try to bring the system to a stabilized frequency. Afterwards, the agent layer will then correct any eventual errors w.r.t the nominal condition through secondary control. The local controller of DG in Control layer includes primary controller and also PI controller in secondary control level.

*1) Primary control*

The primary control, or local control, adjusts the frequency and amplitude of the voltage reference provided to the inner control loop of the voltage source inverter. The droop control method is used to control power sharing between DGs in MG without communication. The main idea of this control level comes from mimicking the self-regulation ability of synchronous generators in power systems as (1), which changes the reference frequency according to the alteration of active power:

$$\begin{cases} f = f_0 - k_P(P - P_0) \\ V = V_0 - k_Q(Q - Q_0) \end{cases} \quad (1)$$

As seen in (1), this control strategy is mainly influenced by the droop coefficients $k_P$ and $k_Q$. From the other terms showing in (1), $f_0$ and $V_0$ are rated frequency and amplitude of grid voltage and $P_0$, $Q_0$ are the normal value of real and reactive power. f and V are the actual measured values of frequency and voltage magnitude when the DG is supplying real power equal to P and reactive power equal to Q.

This control level allows multiple inverter based DGs to share power and maintain the voltage and frequency stability in MG. The frequency and amplitude deviations will be eliminated in secondary control level.

*2) Secondary control*

The secondary control is employed to restore the frequency and voltage to their nominal values after any deviation from these values. This paper will focus only on frequency. The steady-state error is compensated by a PI controller. In particular, the secondary control is computed as

$$\delta_f = K_p \Delta f + K_i \int \Delta f \quad (2)$$

where $K_p$ and $K_i$ are the control parameters of the PI controller, $\Delta f$ is the measured microgrid frequency deviation and $\delta_f$ is the secondary control signal sent to primary control level. $\delta_f$ is then added to the correction given by each loop controller in (1):

$$f = f_0 - k_P(P - P_0) + \delta_f \quad (3)$$

A typical approach is to have a centralized secondary control installed in the MG's central controller which sends the same $\delta_f$ to all local controller units. In our proposal, each local controller is connected to an agent. The controller is required to send its frequency deviation to the corresponding agent and this agent will communicate with others to calculate an average $\Delta f$ from the collected information (either globally or locally with other agents in its neighborhood).

An agent-based consensus algorithm is used in the Agent layer and all agents reach the same value after a specific number of iterations. This is also the average frequency deviation transferred to PI controller. Equation (2) becomes:

$$\delta_f = K_p \Delta f_{average} + K_i \int \Delta f_{average} \quad (4)$$

This control strategy guarantees the operation of MG without the central controller.

*C. Agent layer:*

The Agent layer is a multi-agent system and each agent could get global information by using the consensus algorithm. Agents are put at locations of DG units in the Device layer. The agent network is regulated by the connection ability of DGs. The Agent layer takes the responsibility to send the same signal of frequency deviation to local controllers of DGs in the Control layer. In the centralized control strategy, this role belongs to the microgrid central controller. The instantaneous value of frequency deviations at the output of all DGs are measured in the Device layer and transferred to this layer. The requirement is that the signals are sent to the local controllers at almost the same time and those signals have the same value as in the case of the central controller. These conditions are met through the average consensus process.

The topology of a multi-agent network can be represented as a graph, described by the pair (V, E) where V = {1,…,n} represents the set of vertices (nodes), and E⊆V×V represents the ordered set of edges, or connections, from one node to the other. An edge (i,j)∈ E describes a communication link from node i to node j. The neighbors of node i are denoted by $N_i$={j∈ V:(i,j)∈ E}.

In a network, consensus means to reach convergence regarding a certain quantity of interest that depends on the state of all nodes [9]. A consensus algorithm is an interaction rule that specifies the information exchange between a node and all of its neighbors on the network. The consensus process in a graph network is a set of iterations. Each iteration at one node needs the information from its neighbors and a calculation unit inside for updating the state based on the current state and the collected information. The multi-agent system, which is defined as a set of a number of agents operating in collaboration in order to achieve an overall system-wide goal, is appropriate to be applied for this rule. An agent is set at a network node and combined with others it creates a MAS. The constraint communication in MAS is set by corresponding network topology.

The process will start when a frequency deviation will appear in the MG. The state of the deviation is updated by using the following equation:

$$\Delta f_i[k] = \sum_{j=1}^{n} a_{ij}[k-1] \cdot \Delta f_j[k-1], \quad i = 1, \dots, n \quad (5)$$

where $\Delta f_i[k]$ is the $i^{th}$ node's state at iteration k

$a_{ij}$ is the weight node i assigns to information from node j, as calculated in the adjacency matrix. The elements of the adjacency matrix indicate whether pairs of vertices are adjacent or not in the graph.

$\Delta f_j$ is the state value received from jth DG

The Metropolis Rule [10] in (6) is used to determine the adjacency matrix because it has been shown to guarantee stability, adaptation to topology changes, and near-optimal performance

$$a_{ij} = \begin{cases} \frac{1}{\max(n_i, n_j)}, & i \in N_j\{j\} \\ 1 - \sum_{i \in N_j\{j\}} a_{ij}, & i = j \end{cases} \quad (6)$$

Here, $n_i$ is the number of neighbor nodes of node i.

The consensus process will be converged when $\Delta f_i[k] \rightarrow \Delta f_j[k-1]$ for all i, j = 1,…,n.. The consensus values will be sent to the Control layer in order to regulate the reference frequency. This process finishes only when the MG's frequency is in the operational limit, i.e., the deviation equals to zero.

### III. TEST SETUP

In this paper, a microgrid testing system is setup at the G2elab, Grenoble INP, France. An autonomous microgrid is simulated in real time and connected to hardware agent system to emulate a real communication network for the distributed control method. The test system could be separated into two main parts: real time simulation and communication network.

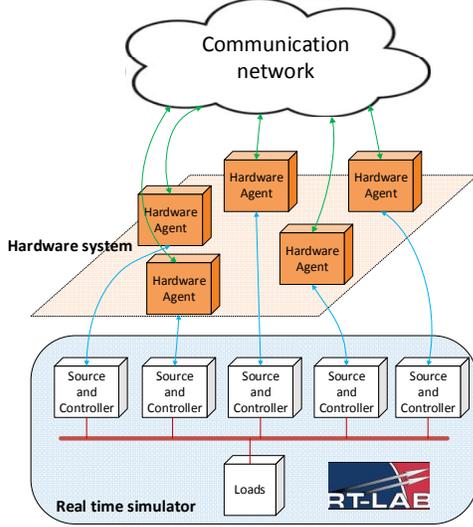

Figure 4. The testing setup

*1) Real time simulation:*

In the test, the simulation in OPAL-RT covers the Device layer and the Control layer in the control structure previously mentioned. To employ the distributed secondary control strategy, an islanded microgrid is simulated in OPAL-RT with five inverter-interfaced DC sources operating in parallel and a variable load. Parameters of system is shown in Table 1.

One of the main advantages of this setup is that it realistically covers the inter-agent communication. The signals between the controllers in the distributed secondary control are not transferred inside OPAL-RT. The controller is connected to a corresponding real hardware agent, which is a ARM-based computer. The hardware agents can send and receive signals to neighbors in the LAN network.

*2) Communication network:*

The angent layer is represented by a distributed computation system consisting of hardware agents and real communication links. Each Agent is a ARM-based computer owning the ability of calculating colective data and connecting to other agents. Five computers corresponding to the five DGs are the nodes in our MAS network. At first, measured frequency deviation signals from the DGs output are sent to the corresponding agent and then transferred to the neighboring agents. The adjacency of our network can be seen in Figure 5. This procedure is executed concurrently in all the nodes of the system. The consensus algorithm is executed by all the hardware agents until they converge to the average value. Finally, these average values from the agent layer are sent back to the controllers running in OPAL-RT, specifically to the PI controllers to compensate the deviation of frequency in MG. If the deviation still exists, the above process will be continued again and it will finish only when the nominal frequency is restored.

To conduct the consensus algorithm in the multi-agent system, a platform called *aiomas* [11] was used. *Aiomas* is an easy-to-use library for request-reply channels, remote procedure calls (RPC) and multi-agent systems. It's written in pure Python on top of *asyncio*. It adds three layers of abstraction around the transports (TCP in this work) that *asyncio* provides. This tool is installed in the ARM-based computers in order to run the consensus process.

TABLE I. PARAMETERS OF COMPONENTS IN MICROGRID

| | Parameter | Value | Unit |
|---|---|---|---|
| **System** | Rated frequency | 50 | Hz |
| **DG1** | Active power set point | $35 \times 10^3$ | W |
| | Droop coefficient | 0.002 | Hz/kW |
| **DG2** | Active power set point | $35 \times 10^3$ | W |
| | Droop coefficient | 0.0022 | Hz/kW |
| **DG3** | Active power set point | $35 \times 10^3$ | W |
| | Droop coefficient | 0.0025 | Hz/kW |
| **DG4** | Active power set point | $35 \times 10^3$ | W |
| | Droop coefficient | 0.0027 | Hz/kW |
| **DG5** | Active power set point | $35 \times 10^3$ | W |
| | Droop coefficient | 0.003 | Hz/kW |

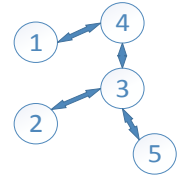

$$E = \begin{bmatrix} 1 & 0 & 0 & 1 & 0 \\ 0 & 1 & 1 & 0 & 0 \\ 0 & 1 & 1 & 1 & 1 \\ 1 & 0 & 1 & 1 & 0 \\ 0 & 0 & 1 & 0 & 1 \end{bmatrix}$$

a) Network topology    b) Adjacency matrix

Figure 5. The topology and adjacency matrix of testing network

The data in the test is transferred between agents by RPC protocol through TCP/IP transport with retransmission ability. The format for serializing and deserializing data is JSON which is a lightweight data interchange format.

## IV. RESULTS AND DISCUSSION

*1) Communication network performance*

We set a Local area network (LAN) in the laboratory which includes five nodes with five ARM-based computers. As mentioned earlier, each computer is an agent that can exchange data with its corresponding lower level controller running in OPAL-RT. The network is configured that one node could only connect with neighboring nodes satisfying network topology in Figure 5. In *aiomas*, the inter-agent message contains a four bytes long header and a payload of arbitrary length as shown in Figure 6. The payload itself is an encoded JSON list, consisting of the message type, a message ID and the actual content. The content here is the called produced in the RPC layer or the data turned back. The latency of data transfer between the *aiomas* agents through the real communication network is depicted in Figure 7 with about 1000 samples. The two numbers on the x-axis indicate the sending and the receiving agent respectively. The variation of time is mainly from ~0.002s to ~0.01s. The time is small, in a range of milliseconds. This is due to the fact that the ARM-based computers are connected to the same local area network. Moreover, there is a small amount of data being transferred. Only the frequency deviation is needed in the consensus process and, as mentioned earlier, it is serialized as JSON, which is considerably lighter than other data formats, e.g, XML. The inter-agent communication might seem fast for a large scale distributed system. However, with the developed system, the quality of service (QoS) of the network can be controlled by adding virtual delays and losses. These aspects however, are beyond the scope of this paper and will be pursued in our future work.

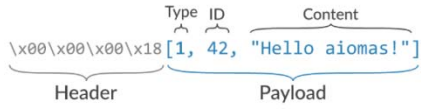

Figure 6. Network message in aiomas [11]

We consider also the consensus processing time (Figure 8), which depends on the topology of network and the data transmission time between nodes. In this case, the average value is reached after 50 iterations. The time in all agents is approximately the same because the computation of each iteration is influenced by the signal received from the neighbors. The average time is at ~0.9s. So the input of secondary control of inverter controller in OPAL-RT is updated after about 0.9s.

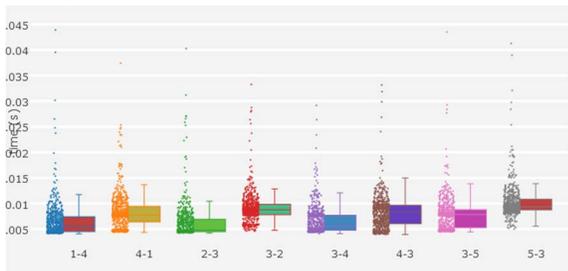

Figure 7. The transmission time in network

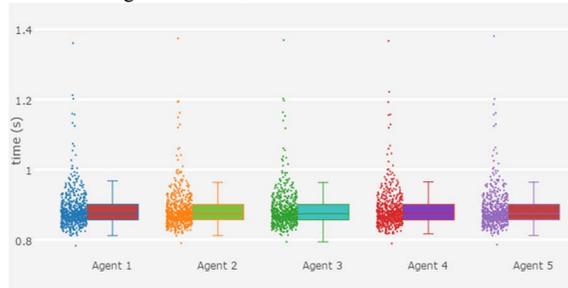

Figure 8. The consensus processing time

*2) Real time simulation performance*

With the system built in the laboratory, the simulation of islanded MG with five inverter-based sources is run in real time in OPAL-RT to evaluate the proposed control strategy. Five ARM-based computers were launched earlier than the starting time of OPAL-RT to be always ready for transferring data and processing the distributed algorithm over the real communication network.

In this test, the active power of the load is increased at 30s and decreased at 60s. Before the load change, the microgrid system was operated in nominal state, meaning that frequency was stable at 50 Hz. The primary control responds rapidly to change the power output of DGs in order to compensate the deficit or excess power in the grid, as shown in Figure 9. The power sharing of the DG is inversely proportional to the droop coefficient value in the controller. The higher the droop factor is, the less power is being generated. It can be observed that in order to keep the frequency steady, this control level reached a new stable state in a very short time-span, due to the electronic based interface and the simple control strategy.

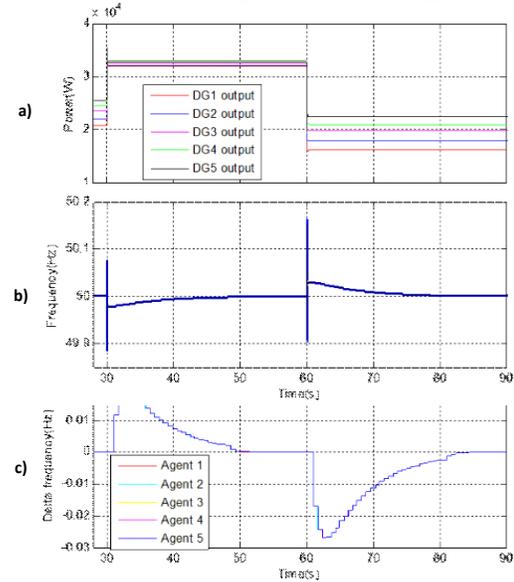

Figure 9. The overal operation of MG from 28s to 90s
a) Output power, b) Frequency, c) The deviation of frequency

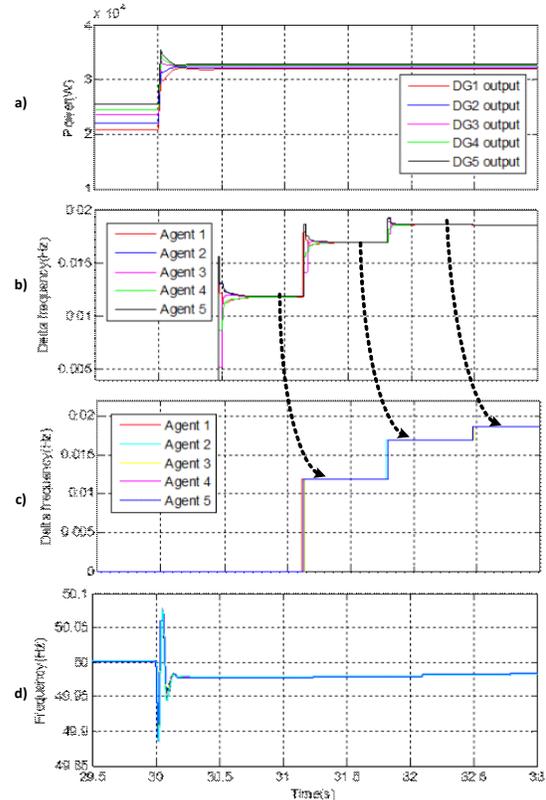

Figure 10. The overal operation of MG from 29.5s to 33s
a) Output power, b) All consensus iteration values of deviation calculated in agents , c) The deviation frequency received from agents, d) Frequency measured at all DGs

The change of DGs active power output ensures the supply-demand balance. Figure 9 shows the overall operation of the system for ~60s. The frequency declines when the load power is higher than the total DGs power output and raises in the reverse process at 30s and 60s respectively. Figure 9.c shows the deviation of frequency signals the OPAL-RT receives from the hardware agents. We can see that the values are almost the same thanks to the synchronous process in all agents, so the controllers of inverters could acquire the proper signals, similar to the centralized control strategy. Consequently, the system is resistant to disturbance and latency communication.

To properly exemplify the operation of the control system, Figure 10 zooms in on Figure 9 for the duration from 29.5s to 33s. Figure 10.b is added to show the result of the calculation for each iteration inside an agent. At first, the primary control keeps the frequency at a stable value, but it is still not yet restored to nominal as the result of the P-f droop. Then, the secondary control starts. After the first consensus process is completed the average values are sent back to primary control. The controllers keep the value transferred from the agents until new converged values are updated. If the frequency deviation still exists, the consensus will go on. Finally, frequency is turned back to nominal state after about 20 seconds.

## V. CONCLUSION

This paper presented a distributed control structure composed of three layers: device layer, control layer and agent layer. Multi-agent system and consensus algorithm are used in the agent layer in order to share the value of deviation frequency. The distributed control strategy reacts to the variation in microgrid frequency in order to keep the system stable without a central controller. This method needs short distance communication, low bandwidth and reliable latency.

An experimental cyber-physical system is built in the laboratory in order to validate the proposed method using a real communication network. The real-time simulation runs in OPAL-RT and it covers the Device and the Control layer. A communication system with five ARM-based computers and a LAN network is set to be in charge of the Agent layer. The results show that this system can properly implement the frequency control without the microgrid central controller. In the future research, the proposed system could be extended to take into account various scenarios at both the physical and the communication layer, e.g., varying communication delays, losses, congestions, etc.